\documentclass[aps,prb,twocolumn,showpacs]{revtex4-1}
\usepackage{graphicx,wrapfig}
\usepackage{fancyhdr}
\usepackage[colorlinks,citecolor = blue,linkcolor=blue]{hyperref}
\usepackage{ifthen}
\usepackage{color}
\usepackage{epsfig}
\usepackage{dcolumn}
\usepackage{float}
\usepackage{mathtools}
\usepackage{comment}
\usepackage{bbm}
\usepackage{amsmath}

\begin{document}

\title{ Electronic structure evolution of the magnetic Weyl semimetal Co$_3$Sn$_2$S$_2$ \\ with hole and electron doping }

\author{Himanshu Lohani}
\affiliation {Universit\'{e} Paris-Saclay, CNRS, Laboratoire de Physique des Solides, 91405, Orsay, France}
\author{Paul Foulquier}
\affiliation {Universit\'{e} Paris-Saclay, CNRS, Laboratoire de Physique des Solides, 91405, Orsay, France}
\author{Patrick Le F\`evre}
\affiliation {Synchrotron SOLEIL, L'Orme des Merisiers, Saint-Aubin-BP 48, 91192 Gif sur Yvette, France}
\author{Fran\c cois Bertran}
\affiliation {Synchrotron SOLEIL, L'Orme des Merisiers, Saint-Aubin-BP 48, 91192 Gif sur Yvette, France}
\author{Doroth\'{e}e Colson}
\affiliation{Universit\'{e} Paris-Saclay, CEA, CNRS, SPEC, 91191, Gif-sur-Yvette, France}
\author{Anne Forget}
\affiliation{Universit\'{e} Paris-Saclay, CEA, CNRS, SPEC, 91191, Gif-sur-Yvette, France}
\author{V\'{e}ronique Brouet}
\affiliation {Universit\'{e} Paris-Saclay, CNRS, Laboratoire de Physique des Solides, 91405, Orsay, France}
\email[]{veronique.brouet@u-psud.fr}

\begin{abstract}
Co$_3$Sn$_2$S$_2$ has been established as a prototype of magnetic Weyl semimetal, exhibiting a ''giant'' anomalous Hall effect in its ferromagnetic phase. An attractive feature of this material is that Weyl points lie close to Fermi level, so that one can expect a high reactivity of the topological properties to hole or electron doping. We present here a direct observation with Angle Resolved Photoemission Spectroscopy of the evolution of the electronic structure under different types of substitutions : In for Sn (hole doping outside the kagome Co plane), Fe for Co (hole doping inside the kagome Co plane) and Ni for Co (electron doping inside the kagome Co plane). We observe clear shifts of selected bands, which are due both to doping and to the reduction of the magnetic splitting by doping. We discriminate between the two by studying the temperature evolution from ferromagnetic to paramagnetic state. We discuss these shifts with the help of DFT calculations using the Virtual Crystal Approximation. We find that these calculations reproduce rather well the evolution with In, but largely fail to capture the effect of Fe and Ni, where local behavior at the impurity site plays an important role.  
\end{abstract}

\maketitle

\section{Introduction}
Topological quantum materials have become a core area of research  in the field of 
condensed matter physics in past decade. Weyl semimetals (WSM) form a part of
  this broad area of research. They feature inverted bands between which a gap closes at only a few non-degenerate Weyl points (WPs). These points act as sources or sinks of Berry flux, giving rise to novel magnetoelectric effects, such as the chiral anomaly\cite{weylrev}. They can be formed either by breaking the inversion symmetry (IS) or the time reversal symmetry (TRS). The best known examples of WSMs are due to IS breaking, as studied in TaS\cite{tas}, TaP\cite{tap} or TaIrTe4\cite{ir}. There are fewer examples of TRS broken WSMs, as they inherently display strong electronic correlations, which are more difficult to control. 
 
 Recently Co$_3$Sn$_2$S$_2$ has been identified as a prototype example of a
 magnetic WSM\cite{ahc}. It has  a rhombohedral structure, where 
 Co atoms form a kagome pattern layer.  Below the Curie temperature 
    of $\sim$ 177 K\cite{ahc}, the Cobalt atoms host local moments M = 0.3 $\mu_B$, which are ferromagnetically aligned along the c-direction\cite{wei} . More complex helimagnetic and antiferromagnetic phases have been suggested to form with doping \cite{heli}. Despite the rather small value of the magnetic moment, a record value of the anomalous Hall conductivity (AHC) has been measured by Liu {\it et al.} \cite{ahc}, which is considered as evidence of the strong Berry curvature around
   the WPs contributing significantly to the  AHC\cite{In,nomura1}.
     Subsequently, Angle-resolved photoelectron spectroscopy (ARPES) has mapped linearly dispersive bulk Weyl bands crossing just above the Fermi level (E$_F$) \cite{yuvil,lei}, in fair agreement with DFT band structure calculations \cite{Sun,nomura}. 
   Moreover, they observed surface states, which could form Fermi arcs between the Weyl points, although there remain discrepancies between the reports about which of these surface states are topological \cite{hasan,yuvil}. Above $T_C$, a $\sim$ 100 meV shift of the bulk bands has been observed \cite{hasan,yuvil1}, with the surface states either vanishing \cite{hasan} or smoothly evolving to Z$_2$ type surface states \cite{yuvil1}.
      
   An attractive feature of Co$_3$Sn$_2$S$_2$ is that WPs lie only $\sim$ 60 meV above E$_F$.
    Therefore, attempts
   have been made to tune E$_F$ position by changing  the chemical potential  by hole
   and electron like doping for fundamental study and applications\cite{hasan1,In,Fe,ni}.
   For doping, Co and Sn are two suitable substitution sites in Co$_3$Sn$_2$S$_2$.
     When Cobalt (atomic number 27) is replaced by Fe (atomic number 26) or Nickel (atomic
    number 28), one can expect a hole and electron like doping, respectively. However, these modifications take place directly in the active kagome Co plane and they may modify more deeply the electronic structure. On the other hand, substitution of In (atomic number 49)  at Sn (atomic number 50) site provides another way to add holes
 in the system. Although there are two Sn sites inside and outside the kagome planes, In was found to substitute preferentially at the interlayer position\cite{jack}.
    
     Earlier experiments have shown that  the T$_c$ of Co$_3$Sn$_2$S$_2$
     decreases by In\cite{In,kasim}, Fe\cite{Fe,Fe1} and Ni\cite{ni,Ni1} dopings. The AHC exhibits a  non-linear change, {\it i. e.} it first increases at
       small doping and then decreases very fast at larger doping. The role of intrinsic \cite{In} and extrinsic \cite{Fe} contributions in these evolutions have been discussed.
Some evidences of a local moment behavior at the dopant site have been reported, either by the observation with STM of a localized bound states at the In site\cite{hasanstm} or by the appearance of a Kondo-like upturn below 50K in the resistivity of Fe-doped samples\cite{Fe}, which is absent for In doping\cite{In}. Therefore, it remains to be understood to which amount the role of substitutions can be regarded as a simple doping effect or a larger perturbation. Will there be a simple rigid-like shift of the electronic structure ? Will it be similar below and above T$_c$ ?

 In this paper, we present direct observation with angle resolved photoelectron spectroscopy (ARPES)
  of the electronic band structure evolution
 of  Co$_3$Sn$_2$S$_2$ doped by In, Fe and Ni. We try to correlate these changes with
 changes in the magnetic moment by also studying the temperature dependence of the electronic structure. We compare our results with those of DFT-based first principles 
     calculations using virtual crystal approximation (VCA) method\cite{vca}  to mimic the doping effect.
     For clarity, we choose significantly doped samples {\it i. e.}  x = 0.44 for 
     In (Co$_3$Sn$_{1.56}$In$_{0.44}$S$_2$), x = 0.42 for Fe (Co$_{2.58}$Fe$_{0.42}$Sn$_2$S$_2$) 
     and x = 0.6 for Ni (Co$_{2.4}$Ni$_{0.6}$Sn$_2$S$_2$) for our study.     
    We find that, as a first approximation, the evolution of the band structure can be described by a rigid-like shift, although the shift is different for Fe and In despite a similar 
     hole doping value. For Ni doping, we observe new, previously unoccupied, electron pockets appearing near $E_F$. With our temperature dependent study, we separate the amount of shift due to doping from the one due to magnetism.

\begin{figure*}
\includegraphics[width=12cm,keepaspectratio]{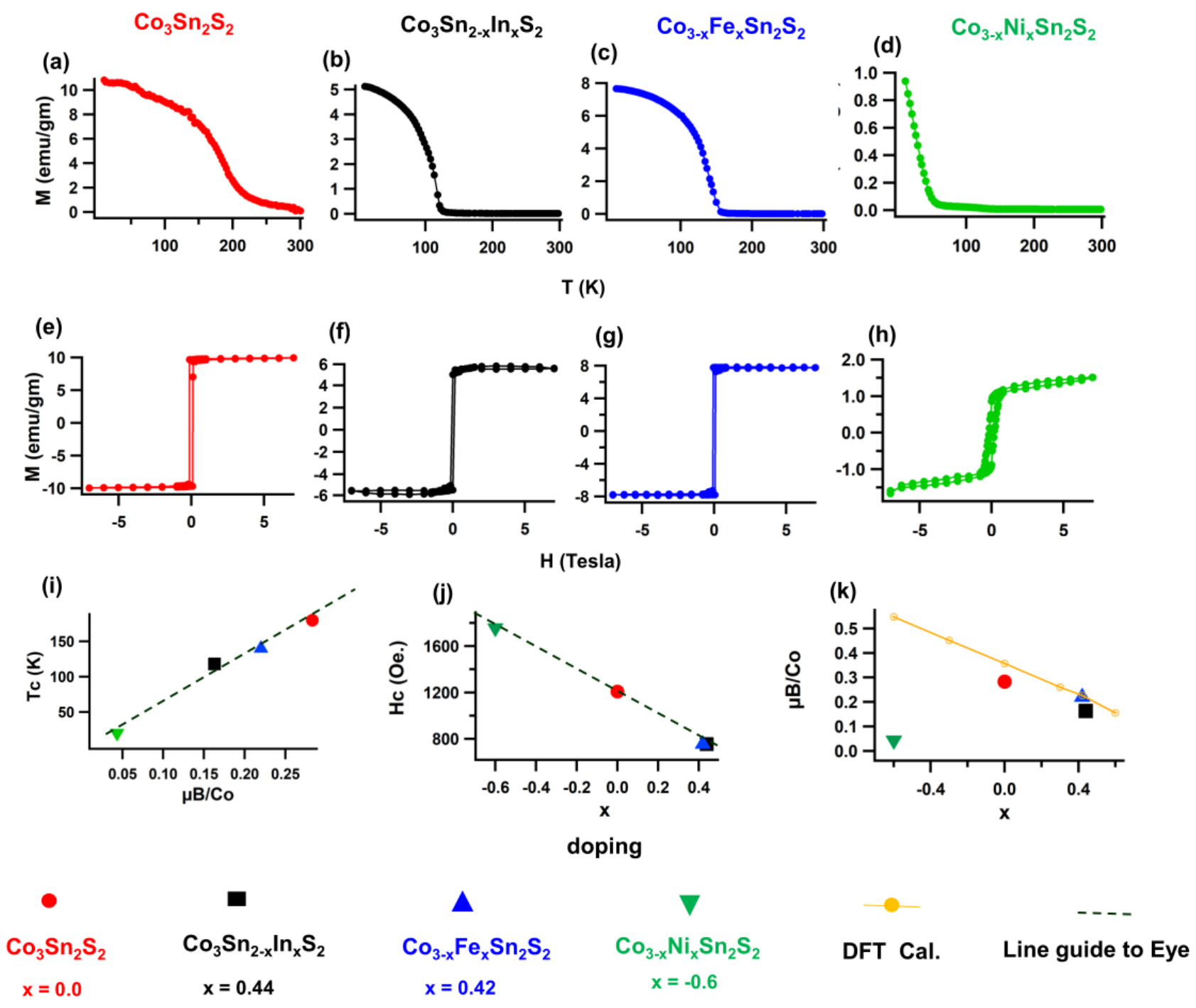}
\caption{\label{fig1}  M vs T plot  at  external field value of 1000 Oe for   Co$_3$Sn$_2$S$_2$ (a)
Co$_3$Sn$_{1.56}$In$_{0.44}$S$_2$ (b),  Co$_{2.58}$Fe$_{0.42}$Sn$_2$S$_2$ (c) and  Co$_{2.4}$Ni$_{0.6}$Sn$_2$S$_2$ (d).   M vs H plot  at temperature  10 K  for   Co$_3$Sn$_2$S$_2$ (e)
Co$_3$Sn$_{1.56}$In$_{0.44}$S$_2$ (f),  Co$_{2.58}$Fe$_{0.42}$Sn$_2$S$_2$ (g) and  Co$_{2.4}$Ni$_{0.6}$Sn$_2$S$_2$ (h).  (i) Curie temperature Vs magnetic moment per Cobalt atom.
 (j) Coercive field as a function of doping concentration x, where
x = 0.0, 0.42, 0.44 and -0.6 corresponds to Co$_3$Sn$_2$S$_2$ (red circle), Co$_3$Sn$_{1.56}$In$_{0.44}$S$_2$ (black square),  Co$_{2.58}$Fe$_{0.42}$Sn$_2$S$_2$ (blue up triangle)  and  Co$_{2.4}$Ni$_{0.6}$Sn$_2$S$_2$ (green down triangle) respectively. (k) Magnetic moment as a function of doping concentration x.
DFT predicted value of magnetic moment of  Co$_3$Sn$_2$S$_2$ for hole (x $>$ 0) and electron (x $<$ 0)
like doping is drawn in (k) by yellow colour line. Black colour dotted line in (i) and (j) is guide to eye 
for the linear decrement. }
\end{figure*}

\section{Experimental  details}
We have grown single crystals of Co$_3$Sn$_2$S$_2$ and  its doped compounds
Co$_3$Sn$_{1.56}$In$_{0.44}$S$_2$,  Co$_{2.58}$Fe$_{0.42}$Sn$_2$S$_2$  and 
 Co$_{2.4}$Ni$_{0.6}$Sn$_2$S$_2$ by using Sn flux method, as reported before \cite{ahc}.
 We checked crystallinity and stoichiometry of the samples  by using XRD
 and energy dispersive X-ray (EDX) measurements.
 Magnetic properties were measured by Quantum Design  SQUID system.
 Photoemission experiments were performed CASSIOPEE-B-ARPES beamline at SOLEIL synchrotron, using SCIENTA-R4000 electron analyser, where a liquid He based cryostat was used to control
 the sample temperature.  The samples were cleaved {\it in-situ} and measured under ultra high vacuum condition
at base pressure $\sim$ 5.0 $\times$ 10$^{-11}$ mbar. The resolution was $\sim$ 15 meV and
0.2$^\circ$ for energy and momentum respectively.

We perform density functional calculations for
Co$_3$Sn$_2$S$_2$ by using the WIEN2K code\cite{wien2k} with the experimentally determined structure of Co$_3$Sn$_2$S$_2$. We estimate the effect of doping using the VCA approximation at the Co site 
and neglecting structural changes.

\section{Magnetic properties}

Fig.\ref{fig1}(a)-(d) shows magnetization as a function of temperature in Co$_3$Sn$_2$S$_2$,  Co$_3$Sn$_{1.56}$In$_{0.44}$S$_2$,  Co$_{2.58}$Fe$_{0.42}$Sn$_2$S$_2$  and  Co$_{2.4}$Ni$_{0.6}$Sn$_2$S$_2$  collected  in field cooled conditions at the external field value of 1000 Oe. 
The external field was parallel to the c-axis of the crystals during the measurements. 
The magnetic moment rises sharply in all the samples indicating their ferromagnetic 
character, which is further confirmed by their ferromagnet hysteresis  loop observed
  in the magnetic moment vs  external field measurement at 10 K shown in Fig.\ref{fig1}(e)-(h).

 We estimate  ferromagnetic transition (T$_c$) by using derivative plot (dM/dT vs T) and
 find that  the T$_c$ of the pristine compound goes down by all types of doping. It is 177K, 140 K,
 118 K and 20 K for pristine, Fe, In and Ni doping. Similarly, the magnetic moment per Cobalt atom obtained from  saturation value of the magnetic moment from the hysteresis curves decreases. 
 It is 0.33, 0.22, 0.15 and 0.04 $\mu_B$ for pristine, Fe, In and Ni doping.
  Fig.\ref{fig1}(i) shows that there is a nearly linear relation between T$_c$ and the saturated magnetic moment.
    On the other hand, the coercive field (H$_c$) decreases by In and Fe doping but it increases in Ni doped 
  samples as depicted in Fig.\ref{fig1}(j).

In a simple reasoning based on the number of available electrons, we expect nearly similar hole dopings for Fe and In (x = 0.42 and x = 0.44, respectively) and an electron doping x = 0.6 for Ni. A change of the total number of electrons can be simulated in a DFT calculation by the VCA method, where the average number of available electrons is assumed for all Co. In Fig.\ref{fig1}(k), we compare  the magnetic moment per cobalt atom calculated
  as a function of doping concentration x by using VCA
  calculation to the experimental values. As was shown in previous reports\cite{nomura1}, for the pristine compound, the DFT calculation captures rather well the experimental moment, although it is slightly overestimated.  Co$_3$Sn$_2$S$_2$ is a half metal, the near E$_F$ DOS is composed of Co-3d states, which acquire
  spin-up polarized character in the ferromagnetic state. In the VCA approach, the moment is simply proportional to the number of electrons filling these states, therefore it increases linearly with the electron filling and keeps increasing from the hole to the electron doped side. However, in the electron doped compound Co$_{2.4}$Ni$_{0.6}$Sn$_2$S$_2$ the magnetic moment is only 17\% of the value of pristine compound, which is quite contrary to this prediction. This clearly shows that this VCA approach is insufficient to treat the effect of doping associated with Co/Ni substitutions. Indeed, DFT calculation using a supercell
approach \cite{ni,Ni1} predict the decrease of magnetic moment, associated with the disorder introduced by Ni. Furthermore, Ni and Co in the kagome plane hold quite different magnetic moments in these calculations, emphasizing that they cannot be treated as a single average atom. 

 Interestingly, the calculated value with VCA for  x = 0.4 (65\% of its value in pristine compound) is rather close to the one found for In doping, where the magnetic moment decreases to  $\sim$ 57\%. This suggests that for substitutions outside the Co network, a simple charge transfer can be assumed. On the contrary, the magnetic moment found  in our experiment for Fe is somewhat higher (77\%), suggesting that doping from Fe and In substitutions are quite different. Earlier report by Kassem {\it et al.}\cite{kasim} have suggested that T$_c$ decreases identically by Fe and In doping, which they considered as a proof of a similar effect of the two types of doping. While the nearly similar change in T$_c$ could be consistent with our findings, we observe that the change in the magnetic moment is
  relatively larger. As doping is an important knob to tune the topological properties of Co$_3$Sn$_2$S$_2$, it is important to understand better how the electronic structure is modified by the different types of substitutions and particularly potential deviations from the rigid band filling picture. We will show that these substitutions indeed have a different impact on the electronic structure. 
	
\begin{figure*}
\includegraphics[width=12cm,keepaspectratio]{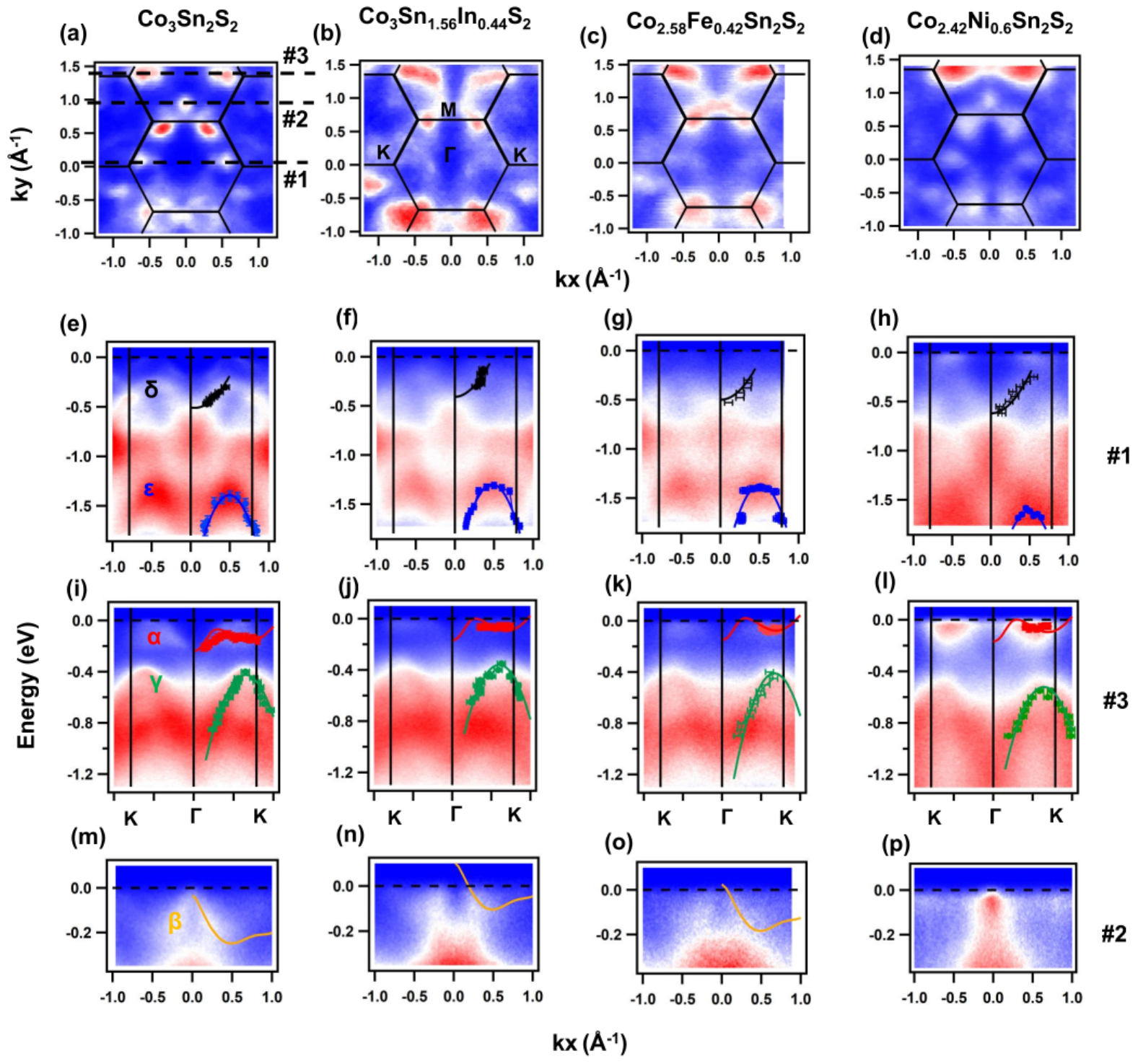}
\caption{\label{fig2} (a-d) FS of Co$_3$Sn$_2$S$_2$,
Co$_3$Sn$_{1.56}$In$_{0.44}$S$_2$, Co$_{2.58}$Fe$_{0.42}$Sn$_2$S$_2$  and Co$_{2.4}$Ni$_{0.6}$Sn$_2$S$_2$, respectively,
collected at 30 K by using 117 eV LH polarized light. Black lines indicate the 2D hexagonal BZ (see Appendix B for more details). 
(e-h), (i-l) and (m-p) ARPES images of the above compounds along the cuts indicated in (a) : cut$\#$1 (k$_y$ $\simeq$0), $\#$3 (k$_y$ $\simeq$1.35) and $\#$2 (k$_y$ $\simeq$ 0.85 \AA{}$^{-1}$), respectively.
$\alpha$, $\beta$, $\gamma$, $\delta$ and $\epsilon$ bands are identified and marked by red, orange, green, black
and blue colours respectively. A parabolic fit is used for the $\gamma$, $\delta$ and $\epsilon$ bands and
a DFT model for $\alpha$ and  $\beta$ (see Appendix B). The DFT calculated bands are renormalized by a factor of 1.4 to match the experimental results. 
  }
\end{figure*}

    \begin{figure*}
\includegraphics[width=12cm,keepaspectratio]{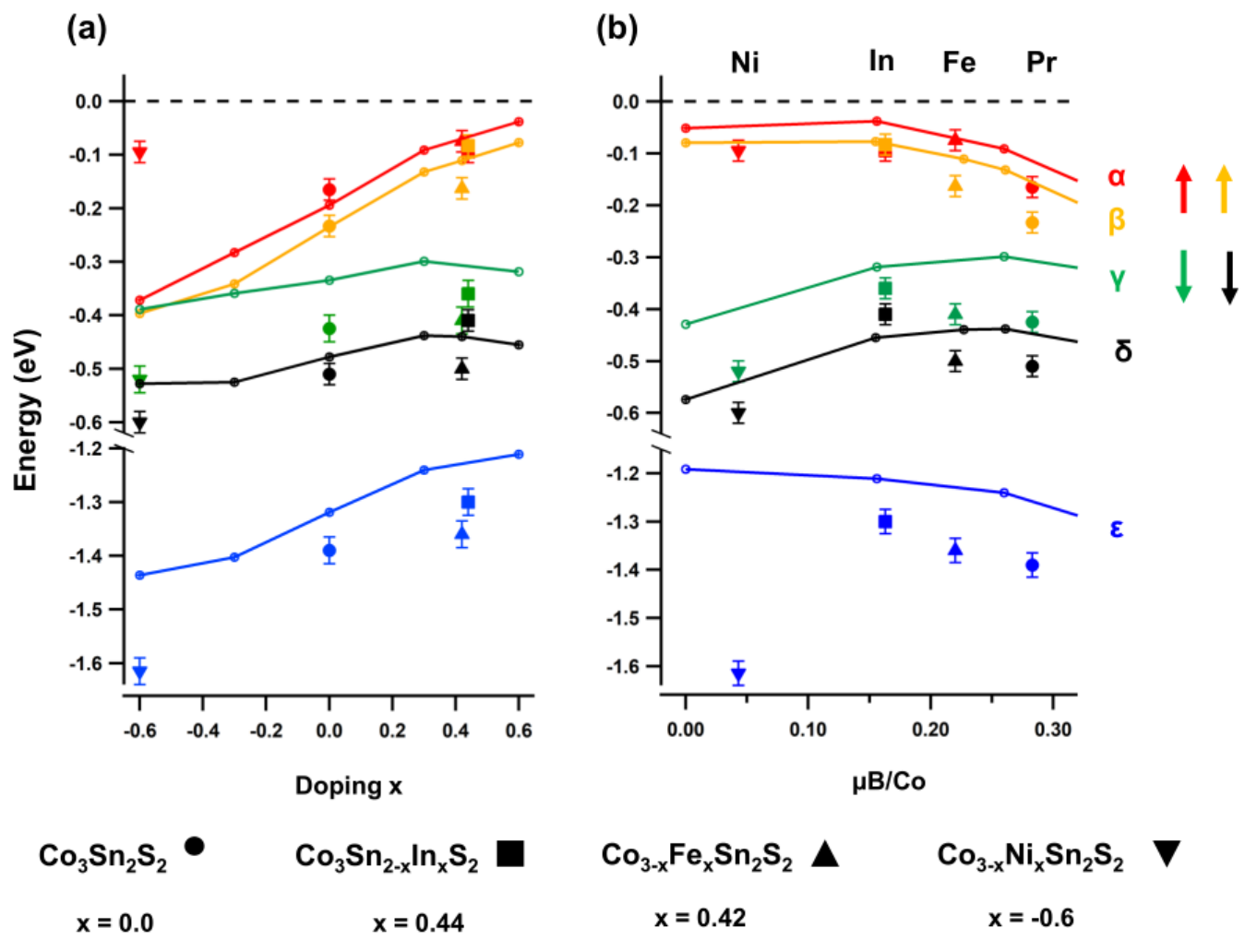}
\caption{\label{fig3}  Position of the $\alpha$ (red), $\beta$ (orange), $\gamma$ (green), $\delta$ (black) and $\epsilon$ (blue) bands defined in Figure 2 with respect to (a) doping (hole (x $>$ 0) and electron (x $<$ 0)) and (b) magnetic moment per Cobalt atom. The solid lines correspond to the DFT predicted doping dependent
change in  the respective bands calculated using the VCA approximation (in (b) the positions for a doping corresponding to the given magnetic moment is used). The position for $\gamma$, $\delta$ and $\epsilon$ are defined at their extrema and for $\alpha$ and $\beta$ at special k-points (see Appendix D).}
\end{figure*}

 \section{Electronic band structure in ferromagnetic state}
We start our investigation by comparing the electronic structure at low temperatures among  the 
four compounds. Fig.\ref{fig2}(a-d) presents Fermi Surface (FS) taken at 30 K by ARPES with a photon energy of 117 eV and linear horizontal (LH) light polarization. In photoemission, the momentum  perpendicular to the in-plane direction (k$_z$) depends on the incident photon energy.  We mapped the k$_z$ dispersion of bands in Co$_3$Sn$_2$S$_2$ (see  appendix A) and find
that this photon energy of 117 eV corresponds to k$_z$ $\simeq$ 0, 
in agreement with previous reports of the literature\cite{yuvil}. 
 In all the FS, a similar pattern can be observed, with high intensity spots along the $\Gamma$-K direction and slightly above M point. 
For clarity, we use here a 2D Brillouin Zone (BZ), more details are given in Appendix B.

 For quantitative comparison of the band structure of the various dopings, we choose three representative cuts in the FSs, through the high symmetry direction  $\Gamma$ - K - M in first BZ (cut$\#$1), through the high intensity point above M at k$_y$ $\simeq$ 0.85 \AA{}$^{-1}$ (cut$\#$2) and through $\Gamma$ - K - M in second BZ (cut $\#$3, this corresponds to k$_z$ = 2/3, see appendix B). In Appendix B, we give for reference the full electronic structure calculated by DFT along each cut. ARPES images taken along these different cuts are presented in Fig.\ref{fig2}. In each case, we identify a similar structure, but with a shift that is doping dependent. Along cut$\#$1 [Fig.\ref{fig2}(e-h)], the band structure in these experimental conditions is characterized by an electron-like band around $\Gamma$ that we call $\delta$ (black line) and an intense parabolic patch at higher binding energy (BE) that we call $\epsilon$ (blue line). Along cut$\#$3 [Fig.\ref{fig2}(i-l)], one band ($\alpha$, red) is identified near E$_F$  and a broad intensity patch ($\gamma$, green)  lies at higher BE. Finally, along cut$\#$2 [Fig.\ref{fig2}(m-o)], a clearly dispersing
band  ($\beta$, orange) is visible and crosses E$_F$ with nearly linear dispersion. As will be justified later, this point is close to the Weyl crossing loop. This
 $\beta$ band is superimposed to an electron like 
band, which is possibly a surface state (it is more clearly visible in circular polarization, see appendix C).

	To track the position of these bands, we extract data points either by fitting the momentum dispersion curves (MDCs) to Lorentzian peaks or by taking the maxima of energy dispersion curves (EDCs) on flat parts of the dispersion where MDC peaks are unresolved. The $\gamma$, $\delta$ and $\epsilon$ points are fitted to  parabolic dispersion to estimate their extreme position. The near E$_F$ bands  $\alpha$ and $\beta$ have more complicated shapes, which are well captured by DFT calculations for the pristine compound after using a renormalization factor of  1.4, as detailed in Appendix B and in agreement with previous reports \cite{yuvil,hasan}. We use these DFT bands as model and shift them to match with the $\alpha$ and $\beta$ bands in doped compounds. For Ni [Fig.\ref{fig2}(p)], it is not possible to describe the $\beta$ band by such a rigid shift. A new electron pocket appear near $E_F$ presumably from a previously unoccupied band. 

 In Fig.\ref{fig3}, we summarize the evolution of the position of each band in the different compounds. We indicate the position of the bottom of the parabola for $\delta$, the top of the parabola for $\gamma$ and $\epsilon$.
  For $\alpha$ and $\beta$ bands, we use the position of the DFT model at k$_x$ = 0.51 \AA{}$^{-1}$
   and k$_x$ = 0.35 \AA{}$^{-1}$ respectively (see Appendix D). 
  In Fig.\ref{fig3}(a), we plot these values as a function of the doping expected in case of complete charge transfer, e.g. a hole concentration with x = 0.44 for Co$_3$Sn$_{1.56}$In$_{0.44}$S$_2$,  x = 0.42 for Co$_{2.58}$Fe$_{0.42}$Sn$_2$S$_2$ and an electron concentration 
  x = -0.6 for Co$_{2.4}$Ni$_{0.6}$Sn$_2$S$_2$. We compare these evolutions with those expected for the DFT bands in the VCA calculation (solid lines). They shift almost linearly with doping, albeit with a different slope for spin-up ($\alpha$, $\beta$ and $\epsilon$) and spin-down ($\gamma$ and $\delta$). The calculated shifts matches rather well with the observed ones on the hole doped side, although it is systematically reduced in Fe (up triangles), compared to In (squares). The position of Fe is usually found half way between those of the pristine compound and the one of In, as if the effective doping was closer to x = 0.2. For Ni, the shift of the spin up band $\alpha$ is opposite to what would be expected for electron doping and strongly deviates from the VCA curve. On the other hand, the position of the spin down bands ($\gamma$ and $\delta$) and $\epsilon$ at high BE are consistent with electron doping and the calculated values, within experimental accuracy. This problem arises because the magnetic splitting is overestimated for Ni by the VCA calculation [see Fig. 1(k)].

	Indeed, as was clear from Fig.\ref{fig1}, doping also modifies the magnetic properties of Co$_3$Sn$_2$S$_2$, which of course will affect the position of the bands in the magnetic state. 
Therefore, we replot in  Fig.\ref{fig3}(b) the position of the bands with respect to magnetic moment, together with the position of the DFT bands at the VCA doping giving rise to this magnetic moment. The DFT predicted spin-up ($\alpha$, $\beta$ and $\epsilon$) and spin-down ($\gamma$ and $\delta$)
   bands move towards the lower
   and higher BE  respectively as magnetic moment  per Cobalt atom decreases. 
   The experimentally observed  position of these  bands do not lie exactly
   on the DFT predicted lines but they show a much better agreement, especially for $\alpha$ in Ni. This means that shift with doping is dominated by the magnetic shift, which was incorrectly predicted for Ni by VCA. 
	On the other hand, the position of $\epsilon$ band in Ni is not well predicted. As this band overlaps with many other bands (see Appendix B), it is difficult to determine whether this difference is significant.

      \begin{figure*}
\includegraphics[width=14cm,keepaspectratio]{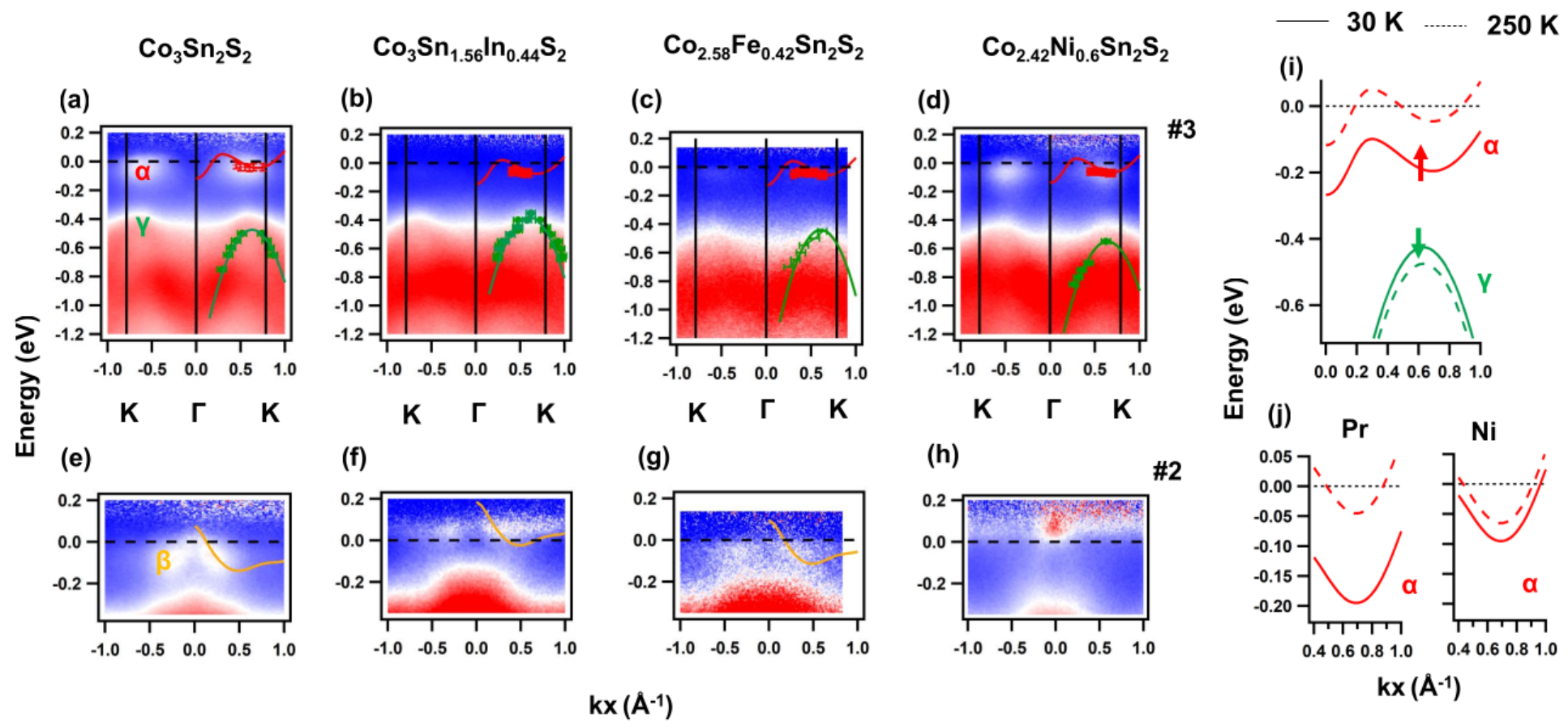}
\caption{\label{fig4} (a-d) ARPES images collected at 250 K along the cut\#3 (as defined in Fig.\ref{fig2}(a)), by using 117 eV LH light
 of Co$_3$Sn$_2$S$_2$, Co$_3$Sn$_{1.56}$In$_{0.44}$S$_2$, Co$_{2.58}$Fe$_{0.42}$Sn$_2$S$_2$
  and Co$_{2.4}$Ni$_{0.6}$Sn$_2$S$_2$ respectively . 
   $\alpha$ and  $\gamma$ bands
  are marked by red and green colours. 
  (e-h) Same along the cut\#2 ($\beta$ band).  
  These images are divided by Fermi-Dirac distribution at 250 K  to highlight the unoccupied region above
  E$_F$.  The overlaid lines are the same as shown in Figure 2 but
  they are shifted to match with the 250 K data.
  (i) Comparison of the models of $\alpha$ and $\gamma$ bands for temperature 30 K (solid line) and 250 K (dotted line). (j) Comparison of the model of $\alpha$ band in Co$_3$Sn$_2$S$_2$ (left) and Co$_{2.4}$Ni$_{0.6}$Sn$_2$S$_2$ (right)
   for temperature 30 K (solid line) and 250 K (dotted line).
  }
\end{figure*}

  \section{Electronic band structure in paramagnetic state}

   Next, we study the evolution with temperature from the ferromagnetic to the paramagnetic state. Fig.\ref{fig4}(a-d) present ARPES images along the cut\#3, in the same conditions as Fig.\ref{fig2}, albeit at 250 K instead of 30 K, where all compounds are in paramagnetic state. In these images, one can notice that the separation becomes larger between the $\gamma$ and $\alpha$ band, spin down and up respectively, as compared to 30 K. In Fig.\ref{fig4}(i), we compare the models observed for $\gamma$ and $\alpha$ bands at 30 K (solid lines)
 and 250 K (dotted lines). When the magnetic splitting closes, the spin up band moves up and the spin down band down, enlarging the gap between them. This larger gap between $\gamma$ and $\alpha$ is then a direct consequence of the magnetic transition. We can also see that at 250 K, when this magnetic shift is removed, the $\alpha$ band in Co$_{2.4}$Ni$_{0.6}$Sn$_2$S$_2$ (Fig.4(d))
lies at higher BE in comparison to its position in Co$_3$Sn$_2$S$_2$ (Fig.4(a)),
   as expected from electron doping. 
	When magnetic shift is present, as sketched in Fig. 4(j), the shift appears different due to the reduced magnetic splitting
   in the Ni doped sample in comparison to the pristine compound.

   \begin{figure*}
\includegraphics[width=13cm,keepaspectratio]{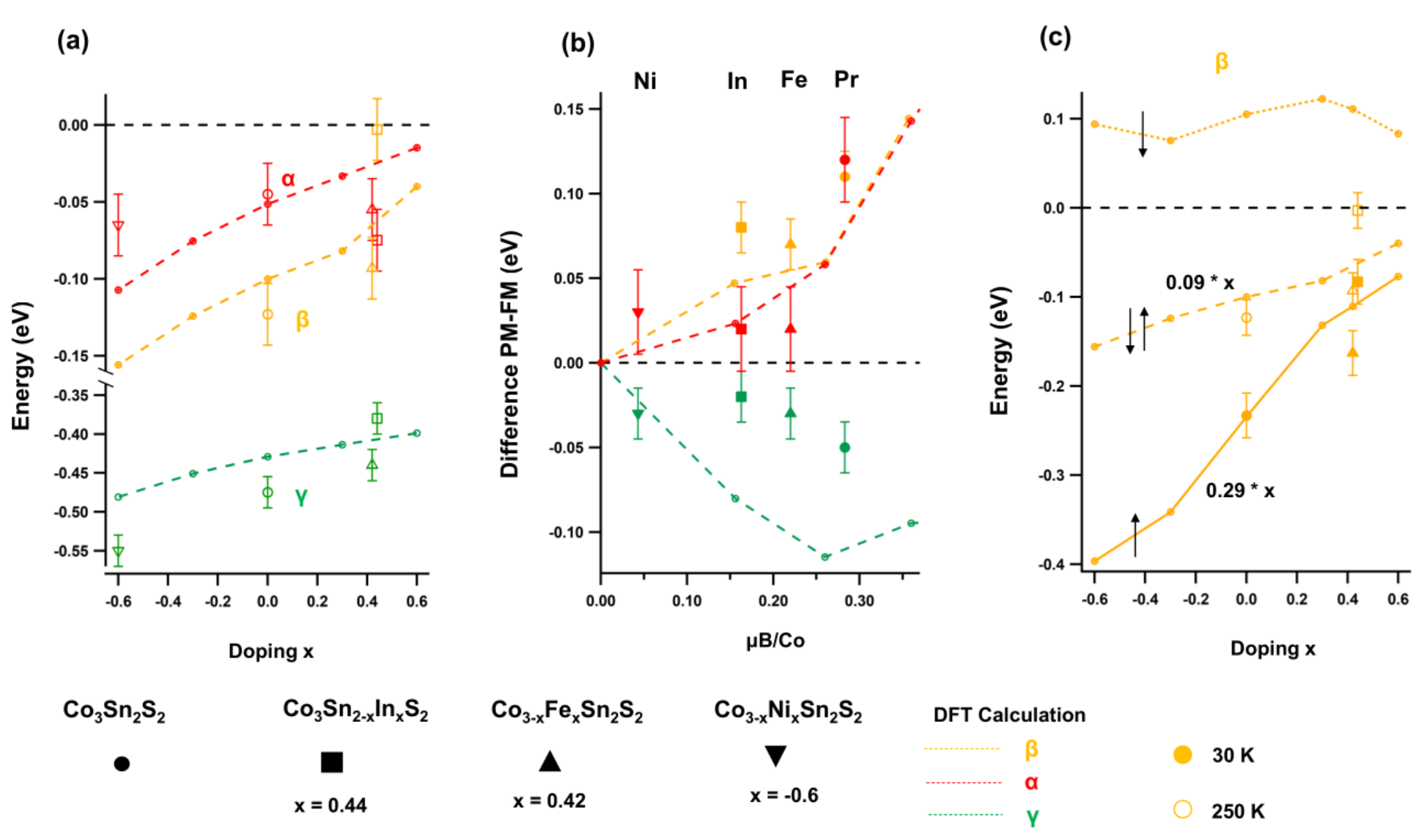}
\caption{\label{fig5} (a) Position of the $\alpha$ (red), $\beta$ (orange) and  $\gamma$ (green) bands
as a function of hole (x $>$ 0) and electron (x $<$ 0) doping at 250 K obtained from the data shown in
Figure 4. Dotted lines represents  their
DFT predicted   value obtained from VCA  calculations  in paramagnetic state. 
(b) Energy difference of the  $\alpha$, $\beta$ and $\gamma$  bands position between paramagnetic state (PM) at  250 K and ferromagnetic state (FM) at 30 K with respect
to the magnetic moment per Cobalt atom obtained from the Fig.\ref{fig1}(k).
(c) $\beta$ band position for different doping, where empty and filled symbols correspond to 
data at 250 K and 30 K respectively.  The center dashed line is the $\beta$ band position for  
 paramagnetic calculation. This splits into spin-up and spin-down bands in ferromagnetic 
 calculations and this spin orientation is
marked by up and down headed black arrows. 
Doping dependent evolution of the   $\beta$ band in PM (0.09) and FM state (0.29)
is obtained by linear fitting  to the paramagnetic
 and spin-up calculations  respectively and written in (c).
 }
\end{figure*}

  \begin{figure*}
\includegraphics[width=15cm,keepaspectratio]{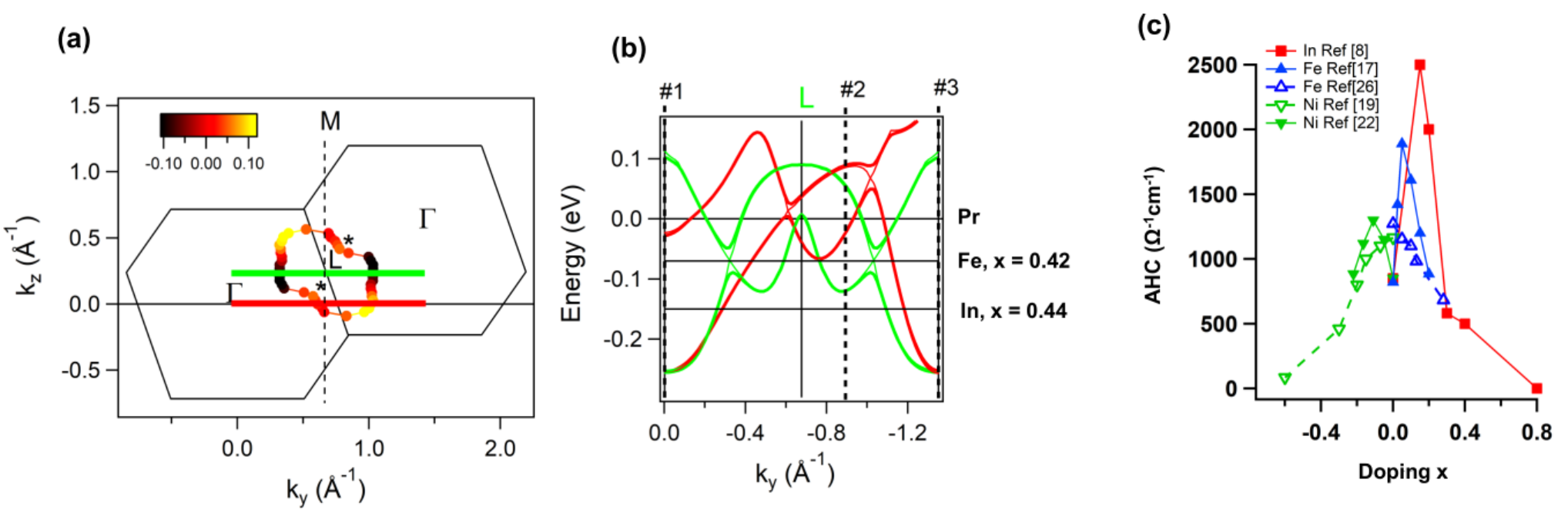}
\caption{\label{fig6} (a) Position of the Weyl loop in the k$_y$ vs k$_z$ plane, as obtained from DFT calculation. The black lines indicate the 3D BZ, the points the position of the crossing between the two bands forming the Weyl loop, the color corresponds to the energy position of the crossing. 
(b) Band dispersion along $\Gamma$ - M at k$_z$ = 0 (red) and k$_z$ =1/3 (green) (thin line without SOC and thick line with SOC). $E_F$ for the Fe and In substitutions   considered in this work at 30 K are indicated. The cuts of Fig. 2 are perpendicular to $k_y$ at positions indicated by the dotted lines. 
(c) AHC as a function of doping reproduced from previous results in the literature for In  Ref[8],
Fe (filled up headed triangle Ref[17] ,  empty up headed triangle Ref[26]) and  Ni (filled down headed triangle Ref[22] ,  empty down headed triangle Ref[19]). 
}
\end{figure*}

 In Fig.\ref{fig4}(e-h), images are shown at 250K along cut\#2. The $\beta$ band similarly exhibits a shift towards E$_F$ as temperature rises to 250 K in the pristine and hole doped compounds (Fig.\ref{fig4}(e-g)). In Ni doped compound (Fig.\ref{fig4}(h)), $\beta$ is still unclear, as it was at 30 K, but the spectral weight we observed at 30K just below  
E$_F$ at k$_x$ = 0.0 \AA{}$^{-1}$ is now shifted above E$_F$. 
This reveals the structure of the bands just above $E_F$ that cannot be observed in the other compounds.

    We summarize in Fig.\ref{fig5}(a) the positions of the $\alpha$, $\beta$ and $\gamma$ bands as a function of doping at 250 K, with
   dotted line representing their values in paramagnetic state obtained from VCA  calculations. The VCA calculation predicts a smooth evolution and this is indeed the qualitative trend. For Ni, there is a clear electron shift for $\gamma$ and $\alpha$ bands. We again find a difference between Fe and In positions, Fe falling roughly between pristine and In. However, the experimental value for $\beta$ found in Co$_3$Sn$_{1.56}$In$_{0.44}$S$_2$ lies clearly above this calculated line, suggesting a deviation for the near E$_F$ states with respect to the calculation. Indeed, even at the VCA level, this part of the band structure is not rigidly shifted (the shift is non-linear for $\beta$ at x=0.6). The shape of the band also slightly changes (see Appendix D). This is probably due to the influence of the hybridized Sn states, and the In doping might amplify the effect. We also note that we neglected here structural evolution, which is not negligible for In \cite{kasim}.
    
We plot in Fig.\ref{fig5}(b) the energy difference between the PM and FM states for the $\alpha$, $\beta$ and $\gamma$  bands 
 with respect to the  magnetic moment. This difference  is positive for the $\alpha$ and $\beta$ bands (spin-up) and negative for the $\gamma$ band (spin-down), as expected from the different spin direction. 
 In addition, the amplitude of the shift is different for the three bands and it shows reasonable
 agreement with VCA calculations.  The larger FM-PM difference (experiment) for $\beta$ band in comparison
  to $\alpha$ and $\gamma$  bands highlights  that Weyl loop is particularly sensitive to magnetic strength 
  compared to the rest of the electronic structure.

To better visualize the relative importance of doping-induced and magnetic-induced shifts, we plot in Fig.\ref{fig5}(c) the calculated and experimental positions for $\beta$ as a function of doping in magnetic and non-magnetic states. The magnetic splitting reduces from 0.35 eV to 0.2 eV between x = 0 and x = 0.4, while $\beta$ shifts towards $E_F$ by 0.05 eV in the PM state and 0.15 eV for the up band in the ferromagnetic state. The magnetic shift is then the dominant effect. In our experiment, the magnetic splitting cannot be determined directly, as the two spin directions of one band are never observed simultaneously. Near E$_F$, one spin direction is unoccupied, above E$_F$, and therefore not observed. At higher BE, the two spin directions overlap too strongly to be resolved. The energy difference between PM state at 250 K and FM state at 30 K of each band gives an indication of the reduction of magnetic splitting, but it also depends on the position of $E_F$ within the band, which depends on details of the semi-metallic structure.

\section{Discussion}
  
  One of the remarkable feature of Co$_3$Sn$_2$S$_2$ is the high value of AHC. It is believed to arise from the proximity of the Weyl points to the Fermi level.
  According to  Kubo formula :
	
   $\sigma_{xy} = \frac{e^2}{h}\sum_{n}\int_{BZ}\frac{d^3 k}{(2\pi)^2}b^{z}_{nk}f(E_{nk}-\mu)$ 
	
	strong  Berry curvature (b$^{z}_{nk}$) around the nodal ring should contribute significantly to the AHC\cite{nomura}.
   Here, $f(E_{nk}-\mu)$ is the Fermi Dirac distribution, where  $\mu$ represents 
 Fermi level, which can be tuned by doping. However, dopants not only modify the AHC by changing the Fermi level (intrinsic factor)  but also
 due to scattering effects (extrinsic factor). The relative magnitude of these two contributions is still debated
 \cite{Fe,Ni1,behnia_ahc}. 
 
 In this situation,  our  doping dependent ARPES study of the electronic structure of Co$_3$Sn$_2$S$_2$ 
 allows to directly locate quantitatively the main bands as a function of doping with respect to the Weyl loop \cite{ahc}. The Weyl loop is formed by the crossing of two inverted bands near E$_F$, the $\beta$ band studied in this paper gives an example of these two bands close to a crossing, which takes place
 just above E$_F$. In Fig.\ref{fig6}(a), we show the position of the loop around L in the k$_y$ - k$_z$ plane, as calculated by DFT similarly to previous studies\cite{yulinsoc,hasan} . The color code indicates the energy position of the crossing with respect to E$_F$. The SOC opens a gap of $\simeq$ 60 meV 
 everywhere\cite{yulinsoc} except at the WPs indicated by asterisks. The band structures calculated at
 k$_z$ = 0 (red line) and k$_z$ = 1/3 (green line) along $k_y$ are indicated in Fig. \ref{fig6}(b), with
  the renormalization of 1.4 indicated by ARPES, with and without SOC (thick and thin lines respectively). The cuts we considered in Fig.\ref{fig2} are perpendicular to this $k_y$ direction at positions indicated on the figure. The typical position of E$_F$ observed for the different dopings are indicated on this plot.  

From this plot, we see that the Fermi level in the pristine compound lies just below the position of the Weyl crossing at k$_z$=0 and above the Weyl crossing at k$_z$ = 1/3. With hole doping, the Fermi level will move to the energy region where the Weyl crossings occurs for k$_z$ = 1/3 rather than around L (green dispersion in Fig.\ref{fig6}). This is where the crossing takes place the lowest energy and where the SOC gap is the largest. It was argued that strong  Berry curvature around the nodal ring should contribute significantly to the AHC in these conditions
\cite{nomura1}. Indeed, previous studies have found that the maximum of AHC occurs for a small hole doping. The previous results by Shen {\it et al.} for Fe \cite{Fe} and Zhou {\it et al.} for In \cite{In} are reproduced in Fig.\ref{fig6}(c). Although the shape of the two hole-doped cases is similar, the peak occurs at different doping values : x = 0.05 in the case of Fe and x = 0.15 in the case of In. From our ARPES study we directly observe
    difference  in the  shifting amount of Weyl band $\beta$ between the 
    In (0.15 eV / 0.12 eV) and Fe (0.07 eV / 0.03 eV) doping at 30 K / 250 K, despite the nearly similar ($\simeq$ 0.4)  doping level. It suggests that the effective doping obtained from Fe is half that of In. If this was the reason for the different maximum of AHC, we would rather expect to find the Fe maximum at twice the value of In. Therefore, the increase of the AHC is probably due to extrinsic contribution,
as was indeed suggested \cite{Fe1}. Fe atoms, going to kagome layer
of Co atoms, enhances asymmetric scattering of the conduction
electrons which is an extrinsic factor to generate AHC \cite{Fe1}
much more efficiently than for In. The maximum value of AHC itself could depend on the details of the electronic structure at one particular doping.

For Ni doping, the AHC measured by Shen {\it et al.}\cite{Ni1}  also increases slightly with small doping (it shows a maximum at x = 0.056) and then decreases. This peak was not as clear in a previous measurement \cite{ni}. This seems to confirm the extrinsic origin of this increase, irrespective of the $E_F$ position.
Recent preprint  has also elaborated that the AHC evolution of Co$_3$Sn$_2$S$_2$ by different (Fe and Ni) dopings depend on  crystal growth method\cite{behnia_ahc}.

\section{Summary}

Our comparative study of  hole 
 (Co$_3$Sn$_{1.56}$In$_{0.44}$S$_2$, Co$_{2.58}$Fe$_{0.42}$Sn$_2$S$_2$)  
 and electron (Co$_{2.4}$Ni$_{0.6}$Sn$_2$S$_2$) doped samples finds that the T$_c$ and magnetic moment of Co$_3$Sn$_2$S$_2$ 
 is suppressed  by all forms of dopings. 
   However, although the number of added holes is nearly the same between the two hole substitutions (In and Fe),
   T$_c$ and magnetic moment evolve differently.
  Furthermore,  the Weyl band $\beta$
     shift towards E$_F$ with respect to pristine  is $\simeq$ 0.15 eV and 0.07 eV respectively
   by In and Fe doping  in ferromagnetic state at 30 K. This shift reduces  to
   $\simeq$  0.12 eV and 0.03 eV in the paramagnetic state   at 250 K.
   These differences show that the dopant sites influences the type of electron transfer, which appears reduced for Fe. In the case of Ni, it is more difficult to evaluate the effective doping, because the $\beta$ band is not clear. On the other hand, our study reveals bands lying close above $E_F$, which are occupied by the added electrons. This complements our knowledge of the near E$_F$ structure of this material. Our investigation confirms that the DFT calculation gives a good starting point to describe the electronic structure, but we also noted some discrepancies about the near E$_F$ states, for example the position of $\beta$ in the PM state for In or of the pockets above E$_F$ revealed by Ni doping. Although rather small, these deviations can have significant consequences for the shallow pockets governing the properties in this semi-metallic structure. 
	
	 In this ARPES investigation, we have given particular attention to representative bands of the spin up and spin down directions. This has clarified how doping and magnetism influence their relative positions.  We find a shift of $\simeq$ 0.29*x eV in the magnetic state for states near E$_F$ and $\simeq$ 0.09*x eV in the paramagnetic state (here, x is the doping concentration). This complements the information already published on the pristine compound 
\cite{yuvil,lei,hasan,yuvil1,yulinsoc,vishik}, which focused essentially on the spin up near E$_F$ states. It is quite remarkable that DFT predictions describe rather well the magnetic transition. This contrasts with the situation in other ferromagnetic systems, like Fe$_3$GeTe$_2$, where the magnetic splitting does not vanish clearly at T$_c$\cite{fegete,fegete1}. This suggests an itinerant origin of magnetism with relatively small correlation effects in Co$_3$Sn$_2$S$_2$.

\section{Acknowledgment}This work benefited from financial support by Investissement d’Avenir LabEx PALM (Grant No.ANR-10-LABX-0039-PALM).

\newpage
\section{ Appendix}
      
\subsection{ Estimation of the value of $k_z$ at 117eV}

\begin{figure*}
\includegraphics[width=12cm,keepaspectratio]{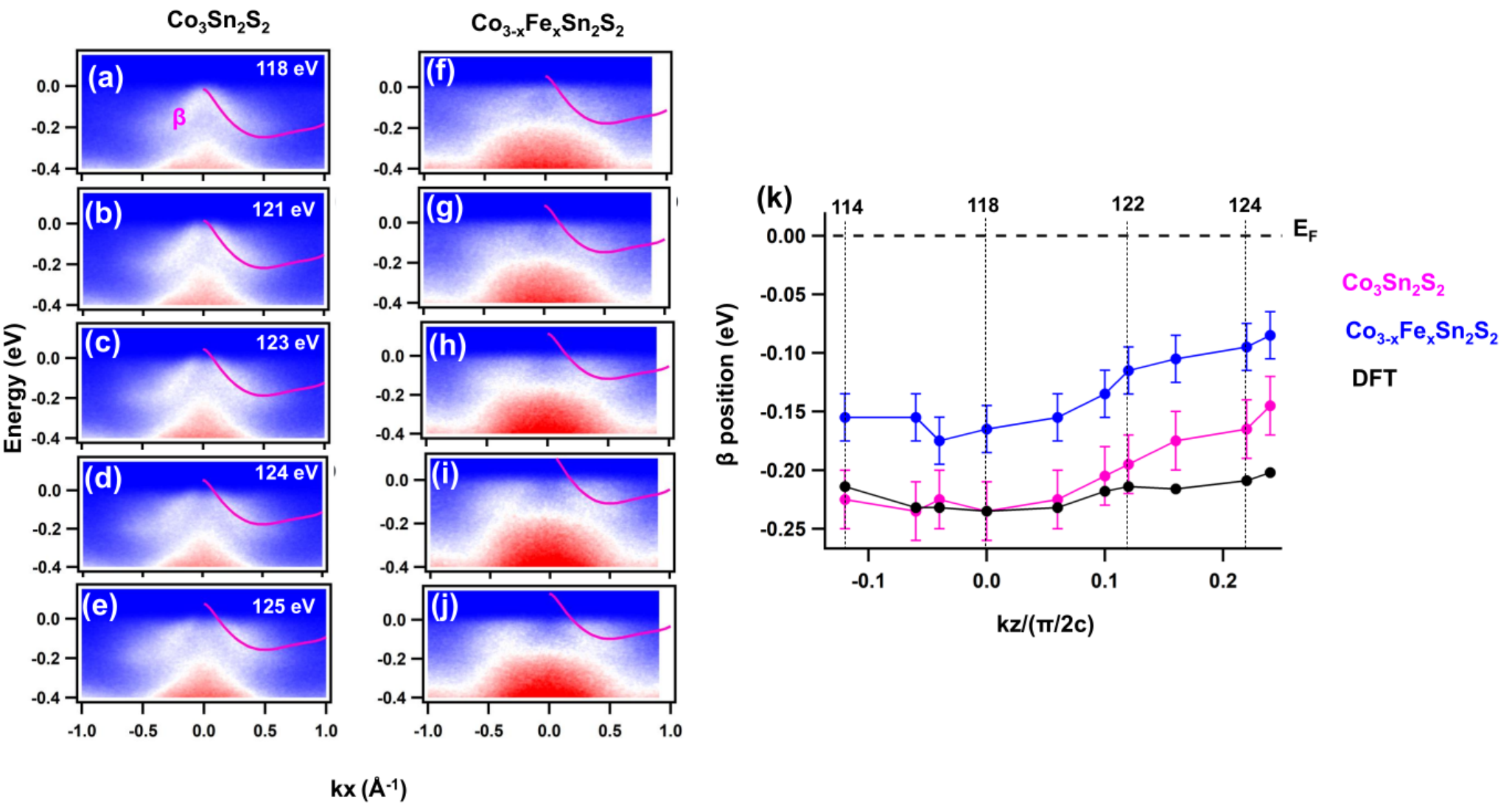}
\caption{\label{fig8} (a/f), (b/g), (c/h), (d/i), (e/j) ARPES images displaying the $\beta$ band 
dispersion for LH polarization at 30 K and photon energies  118 eV, 121 eV,
123 eV, 124 eV  and 125 eV  for Co$_3$Sn$_2$S$_2$ and 
Co$_{2.58}$Fe$_{0.42}$Sn$_2$S$_2$ respectively, where DFT predicted $\beta$ band (magenta colour)
is overlaid in these images. (k) Bottom part of  the $\beta$ band position observed in Co$_3$Sn$_2$S$_2$ (magenta) and Co$_{2.58}$Fe$_{0.42}$Sn$_2$S$_2$ (blue) as a function of k$_z$ normalized by $\pi$/2c factor, where c is the distance between planes in Co$_3$Sn$_2$S$_2$. The black data points represent  DFT predicted  actual k$_z$ dispersion of the $\beta$ band in pristine compound.
}
\end{figure*}

       In photoemission, the dispersion perpendicular to the plane direction (k$_z$) can be mapped by varying the photon energy\cite{huf}. One can convert the photon energy to the k$_z$ value by using formula 
 $k_z$ = $\sqrt{\frac{2m}{\hbar^2}}\sqrt{(h\nu-\phi-E_B)cos^2\theta+V_0}$, where $\phi$, E$_B$, $\theta$ and V$_0$ correspond to work function, BE of photoelectron, emission angle of photoelectron with respect to the sample normal and  inner potential of the sample respectively. V$_0$ is estimated from the periodicity of the observed dispersion. We found that the value of V$_0$=11eV allows a good description of the experimental periodicity, which is in agreement with earlier works\cite{yuvil}. 
  Fig.\ref{fig8}(a)-(e) show the $\beta$ band dispersion for the photon energy range 118 eV - 125 eV at 30 K. We find that the $\beta$ band position is minimum near 117eV and shifts towards $E_F$ with increasing photon energy. As it does not drastically changes its shape, we follow its position by shifting the DFT dispersion at k$_z$=0 (after renormalization by a factor of 1.4, see next section). This behavior, reported in Fig. \ref{fig8}(k) is in reasonably good correspondence with the DFT calculation for the $\beta$ band. This allows to assign a value k$_z$=0 to our experiment at 117eV. Note that the well defined k$_z$ variation of the $\beta$ band implies it is a bulk band.

\subsection{Comparison to band structure calculations}
\begin{figure*}
\includegraphics[width=14cm,keepaspectratio]{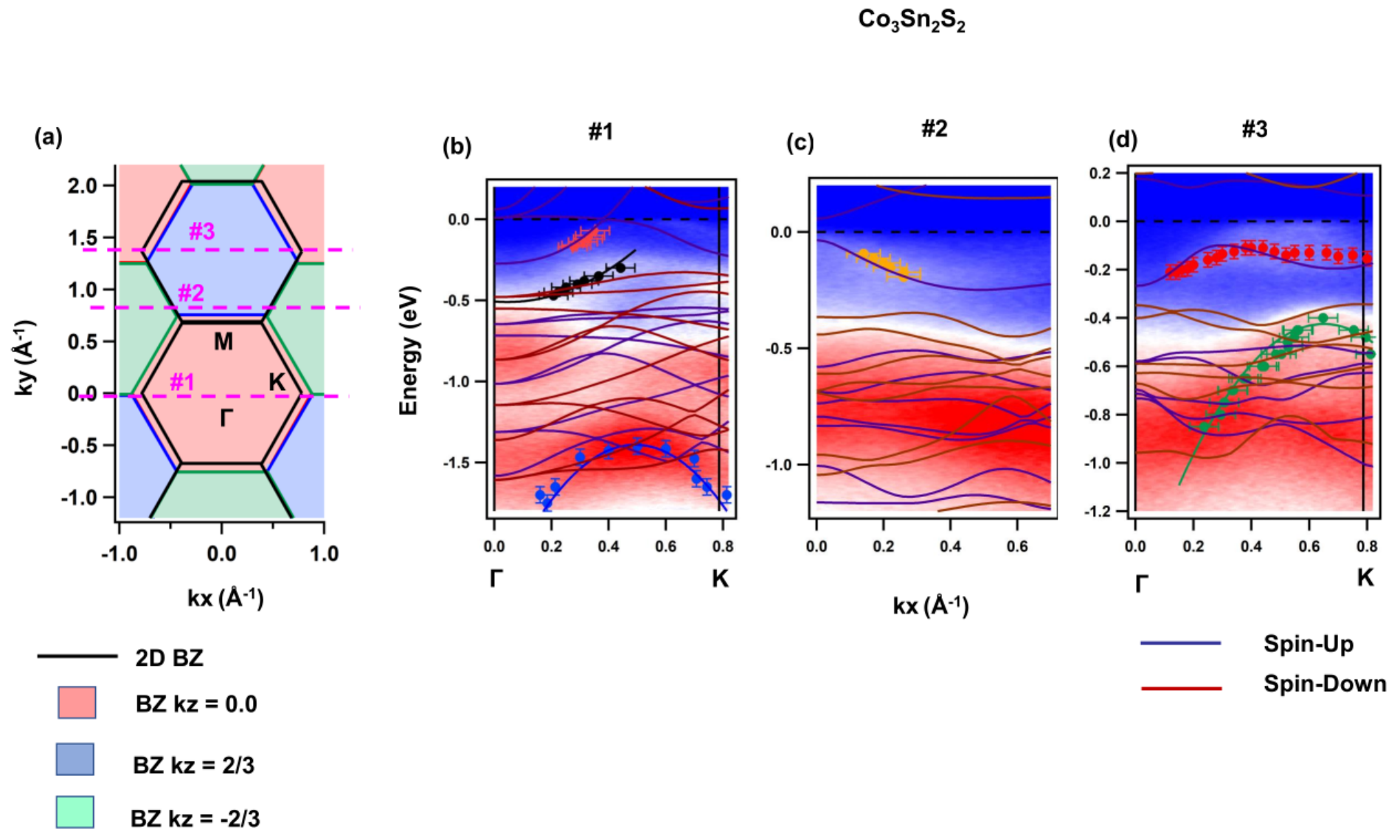}
\caption{\label{fig9}(a) BZ boundary of  Co$_3$Sn$_2$S$_2$ in k$_x$ and k$_y$ plane,
where red, blue and green colours filled area correspond to the BZ boundary 
k$_z$ value = 0.0, 2/3$\pi$/c and -2/3$\pi$/c  respectively. 
(b), (c) and (d)  ARPES images of Co$_3$Sn$_2$S$_2$ taken at 30 K  using
     LH polarized light of 117 eV along the cut \#1, \#2 and \#3 respectively.
 Violet and marron colours   represent  spin-up and spin-down bands respectively
 obtained from the  DFT calculation  for ferromagnetic phase along the respective cuts.
  These bands  are renormalized by a factor of 1.4 in order to match with experimental bands. }
\end{figure*}

In this section, we compare the bands $\alpha$, $\beta$, $\gamma$, $\delta$ and $\epsilon$ described in the text with respect to the DFT calculation. 
     Fig.\ref{fig9}(a) displays the BZ boundaries expected when the normal incidence corresponds to k$_z$=0 ($\Gamma$ point). 
		The black hexagon corresponds to the 2D BZ of the kagome plane, with $\Gamma$K and $\Gamma$M as high symmetry directions. In color, we represent the cut of the 3D BZ at different k$_z$ value, red for k$_z$ = 0.0, blue for k$_z$ = 2/3$\pi$/c and green for k$_z$ = -2/3$\pi$/c. Indeed, adjacent BZs have different k$_z$ value with respect to central BZ in the rhomboedral structure  (see stacking in Fig. 6). 
        The three cuts we consider in the main text are indicated and the ARPES intensity plots taken for the pristine compound at 30 K using
     LH polarized light of 117 eV along these cuts are presented in Fig.\ref{fig9}(b), (c) and (d), respectively. The points extracted from the images are indicated. DFT calculated bands for ferromagnetic phase (normalized by a factor of 1.4) 
     are superimposed in these images, where violet and marron colours  represent 
     spin-up and spin-down bands respectively. Along the cut \#1 
      the $\alpha$ (red) and $\delta$ (black) bands are well reconciled with the near
       E$_F$ spin-up and spin-down bands 
 respectively. Similarly, the $\beta$ (orange) and $\alpha$ bands along the cut \#2 and \#3 exhibit
  a reasonable matching with the DFT prediction. The intensity  patch  of  $\epsilon$ (blue) band  
 falls in an energy window where more than one band are present.

\subsection{Surface state behind the $\beta$ band}

\begin{figure*}
\includegraphics[width=10cm,keepaspectratio]{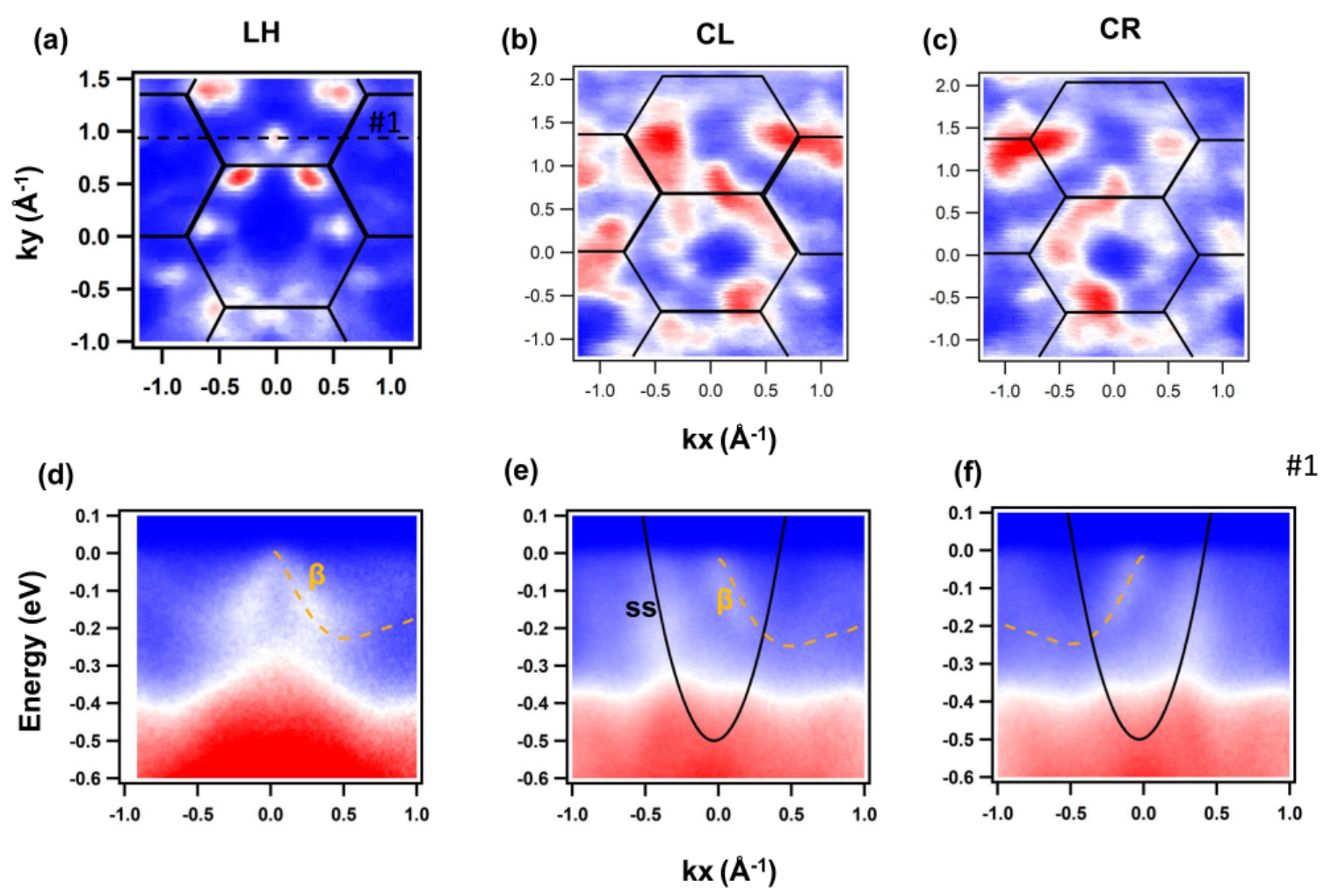}
\caption{\label{fig10} Fermi Surface  of Co$_3$Sn$_2$S$_2$ taken at 30 K by using
LH (a), CL (b) and CR (c) polarized light of photon energy 117 eV. 
(d), (e) and (f) ARPES images along cut\#1 taken from (a), (b) and (c) respectively.
 }
\end{figure*}

Fig.\ref{fig10}(a), (b) and (c) show FS  of Co$_3$Sn$_2$S$_2$ taken at 30 K by using
linear horizontal (LH), circular left (CL) and circular right (CR) 
 polarized light of photon energy 117 eV. 
  ARPES intensity plots along cut\#1 are displayed in 
  Fig.\ref{fig10}(d), (e) and (f) respectively.
  In LH polarization mainly the $\beta$ band is visible but in circular polarization we can see
  another electron like band (SS) is present just behind the $\beta$ band.  No such electron like band is
  present in bulk DFT calculation along this cut, so most possibly  this is a surface related band.
  The $\beta$  and SS band show opposite response to circularly polarized light, {\it i. e.} left and right branch
  of the $\beta$ and SS are more intense in CL light and vice versa in CR light. 
  
\subsection{Determination of $\alpha$ and  $\beta$ band position}

\begin{figure*}
\includegraphics[width=12cm,keepaspectratio]{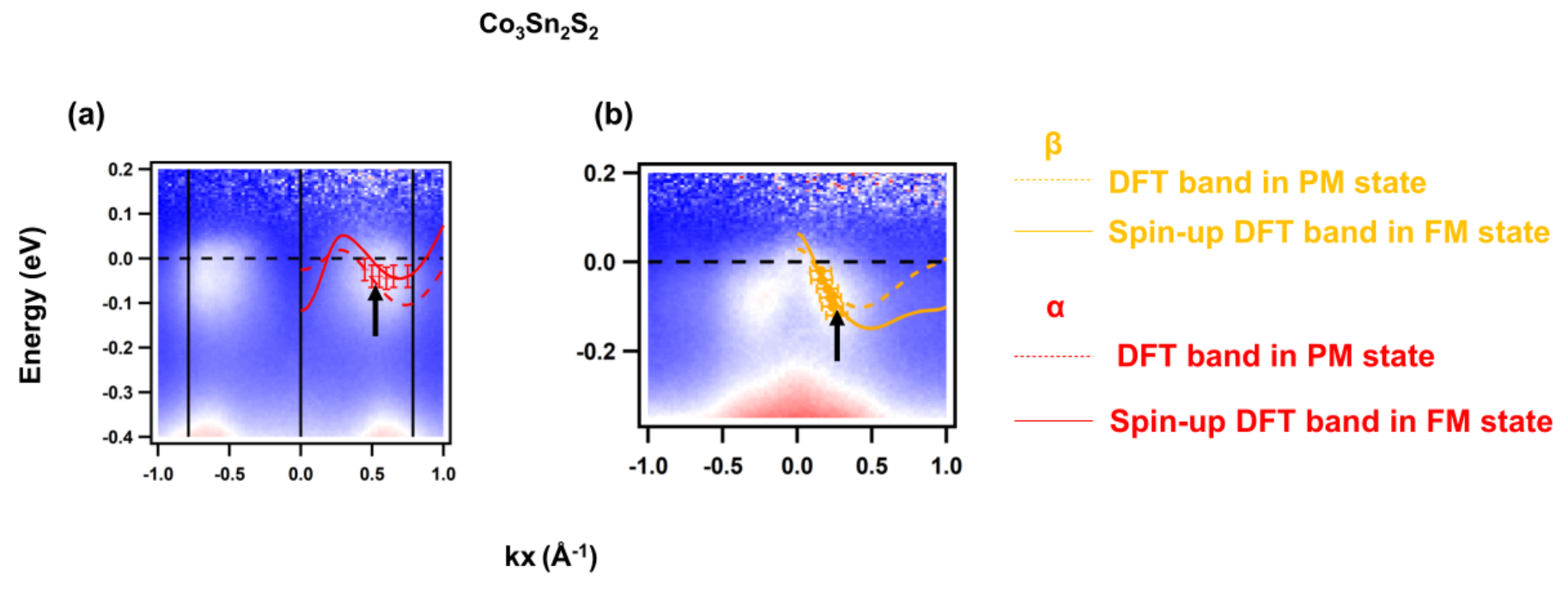}
\caption{\label{fig11}   ARPES image of Co$_3$Sn$_2$S$_2$  
along the cut\#3 (a) and \#2 (b)  taken at 250 K by using LH polarized light of 117 eV.  
These images are divided by Fermi-Dirac distribution at 250 K  to highlight the unoccupied region above
  E$_F$.  In these images red and orange solid and dotted lines represent
  DFT predicted $\alpha$ and $\beta$ bands (renormalized by 1.4)
calculated for FM and PM state respectively. Black arrow indicates
  k-position of the $\alpha$/$\beta$ band which is used to estimate their energy value for 
  experimental  (from FM DFT  band) and theoretical  (from PM DFT  band) value to plot
  in Fig.\ref{fig5}(a). }
\end{figure*}

To determine the position of bands for Fig. 3 and 5, we use the position at k$_x$ = 0.51 \AA{}$^{-1}$
 for $\alpha$ and k$_x$ = 0.35 \AA{}$^{-1}$  for $\beta$, both for experiment and calculation. These k-points are shown by black arrows on \ref{fig11}(a) and (b) for images taken for Co$_3$Sn$_2$S$_2$ along the cut\#3 and \#2 at 250 K. 

The DFT predicted bands, renormalized by 1.4, are shown for the PM state (dashed line) and for the FM state (solid line, spin up) with an up-shift needed to reach the experimental position.

In Fig. 4, to describe the $\alpha$ and $\beta$ bands at 
 250 K, we shift the DFT predicted band calculated for ferromagnetic state (renormalized by 1.4) towards E$_F$ to
 match with the ARPES data at 250 K.
 Thus, we use the same model to estimate the $\alpha$ band position at low T (30 K) and high T (250 K) and we can define a shift between them, independently of details of the calculation.

 However, we can notice there is a slight difference in the DFT calculated band shape in FM and PM state. This shows the limit of the rigid shift approach. The sensitivity of the near $E_F$ structures beyond a simple shift can play an important role for this semi-metallic state with few carriers, but they are beyond the scope of this report.

\end{document}